\documentclass[aps,prb,showpacs,twocolumn,reprint,superscriptaddress]{revtex4-1}
\usepackage{amsmath}
\usepackage{amssymb}
\usepackage{graphicx}
\usepackage{color}
\usepackage{textcomp}
\begin{document}
\title{Designing substrates for silicene and germanene: First-principles calculations}
\author {M. X. Chen}
\email{chen59@uwm.edu}
\affiliation{Department of Physics, University of Wisconsin, Milwaukee, Wisconsin 53211, USA}
\author {Z. Zhong}
\affiliation{Max-Planck-Institute for solid state research, Heisenbergstrasse 1, 70569 Stuttgart, Germany}
\author {M. Weinert}
\affiliation{Department of Physics, University of Wisconsin, Milwaukee, Wisconsin 53211, USA}

\date{\today}

\begin{abstract}
We propose a guideline for exploring substrates that stabilize the monolayer honeycomb structure of silicene and germanene  
while simultaneously preserve the Dirac states: in addition to have a strong binding energy to the monolayer,
a suitable substrate should be a large-gap semiconductor with a proper workfunction
such that the Dirac point lies in the gap and far from the substrate states when their bands align.
We illustrate our idea by performing first-principles calculations for silicene and 
germanene on the Al-terminated (0001) surface of Al$_2$O$_3$.
The overlaid monolayers on Al-terminated Al$_2$O$_3$(0001) retain 
the main structural profile of the low-buckled honeycomb structure 
via a binding energy comparable to the one between silicene and Ag(111). 
Unfolded band structure derived from the $k$-projection method reveals that 
gapped Dirac cone is formed at the K point due to the structural distortion and the interaction with the substrate.
The gaps of 0.4 eV and 0.3 eV respectively for the supported silicene and germanene 
suggest that they may have potential applications in nanoelectronics.
\end{abstract}

\pacs{71.20.-b,73.20.-r,73.22.Pr}

\maketitle
Silicene, germanene and stanene have attracted considerable attention in recent years since the discovery of graphene 
for they can host a variety of novel physical phenomena, such as 
quantum spin Hall effect \cite{liu_quantum_2011,xu_large-gap_2013,C2JM30960B,PhysRevB.89.195403}, 
quantum anomalous Hall effect \cite{valley-polarized_2012,jp5058644,PhysRevB.89.035409,PhysRevB.89.195444}
and valley Hall effect\cite{PhysRevB.87.155415,PhysRevB.87.235426},
which are related to the Dirac electrons in these unique two-dimensional systems.
Moreover, they potentially exhibit extraordinary electronic properties\cite{ni_tunable_2012,drummond_electrically_2012,Ezawa_NJP_2012},
holding significant promise for nanoelectronic devices.

However, in sharp contrast to graphene which can be mechanically exfoliated from graphite
silicene, germanene and stanene have to be grown on substrates.
Up to now, silicene monolayers and multi-layers have been obtained 
by molecular-beam epitaxial (MBE) growth on Ag(111)
\cite{lalmi_epitaxial_2010,vogt_silicene:_2012,chen_evidence_2012,feng_evidence_2012,arafune_structural_2013,chen_spontaneous_2013,lin_substrate-induced_2013,
liu_various_2014,de_padova_evidence_2013,mannix_silicon_2014},
Ir(111)\cite{meng_buckled_2013},
and ZrB$_2$(0001)\cite{fleurence_experimental_2012}.
These systems exhibit a variety of structural reconstructions  
such as (4$\times$4) \cite{vogt_silicene:_2012}, ($\sqrt{12}\times\sqrt{12}$), 
($\sqrt{13}\times\sqrt{13}$) with respect to the unit cell of Ag(111) and 
($\sqrt{3}\times\sqrt{3}$) \cite{chen_evidence_2012} with respect to the 1 $\times$ 1 silicene.
Most recently, germanene
\cite{li_buckled_2014,davila_germanene:_2014,derivaz_continuous_2015}
and stanene\cite{zhu_epitaxial_2015} were also obtained by MBE growth.
However, the structural reconstruction and the strong hybridization between the grown monolayers and the substrates
may significantly modify the electronic structures of the overlayers 
\cite{lin_substrate-induced_2013,chen_revealing_2014,guo_absence_2013,wang_absence_2013,
cahangirov_electronic_2013,quhe_does_2014,mahatha_silicene_2014}.
 Although theoretical calculations indicate that the Dirac states in the monolayers
can be well preserved when they interact with the substrate via vdW-type interactions
\cite{guo_absence_2013,cai_stability_2013,Houssa_APL_2010,Kaloni_JAP_2013,Ding_APL_2013,Scalise_2D_2014,Kokott_JPCM_2014,C4CP00089G},
the weak layer interactions are unfavorable for stabilizing the low-buckled honeycomb structure.
On the other band, it is predicted that the monolayer structure can be stabilized on semiconducting surfaces
such as the (111) surface of GaP, GaAs, ZnS, and ZnSe \cite{Bhattacharya_APL_2013,PSSR:PSSR201510338}. 
However, electronic bands lose features of the Dirac states.
Therefore, it is essentially important to explore substrates that stabilize the monolayer honeycomb structure  
while simultaneously preserve the Dirac states in the silicene/germanene monolayer.

we propose that a suitable substrate for silicene and germanene should satisfy the following criteria:
(i) it has a strong binding energy to the monolayer;
(ii) it has a large energy gap;
(iii) it has a proper workfunction such that the Dirac point of the monolayer 
lies in the gap and far from the substrate states when their bands align.
We illustrate our idea by performing first-principles calculations for silicene and 
germanene on the Al-terminated (0001) surface of Al$_2$O$_3$.
Despite the binding energy between the monolayer structure and Al$_2$O$_3$ comparable to that for metal-supported systems,
unfolded bands reveal that the Dirac states in silicene and germanene are basically preserved on Al-terminated Al$_2$O$_3$(0001).
The structural distortion along with the interaction from the substrate induce a gap opening at K, 
which makes them promising for potential applications in nanoelectronic devices.
Moreover, our results reveal that the interaction from the substrate gives rise to minigaps 
in the band structure of the overlaid monolayers, 
which will lead to Fermi velocity renormalization and thus affect the electronic transistor properties.

\begin{figure}
  \includegraphics[width=0.48\textwidth]{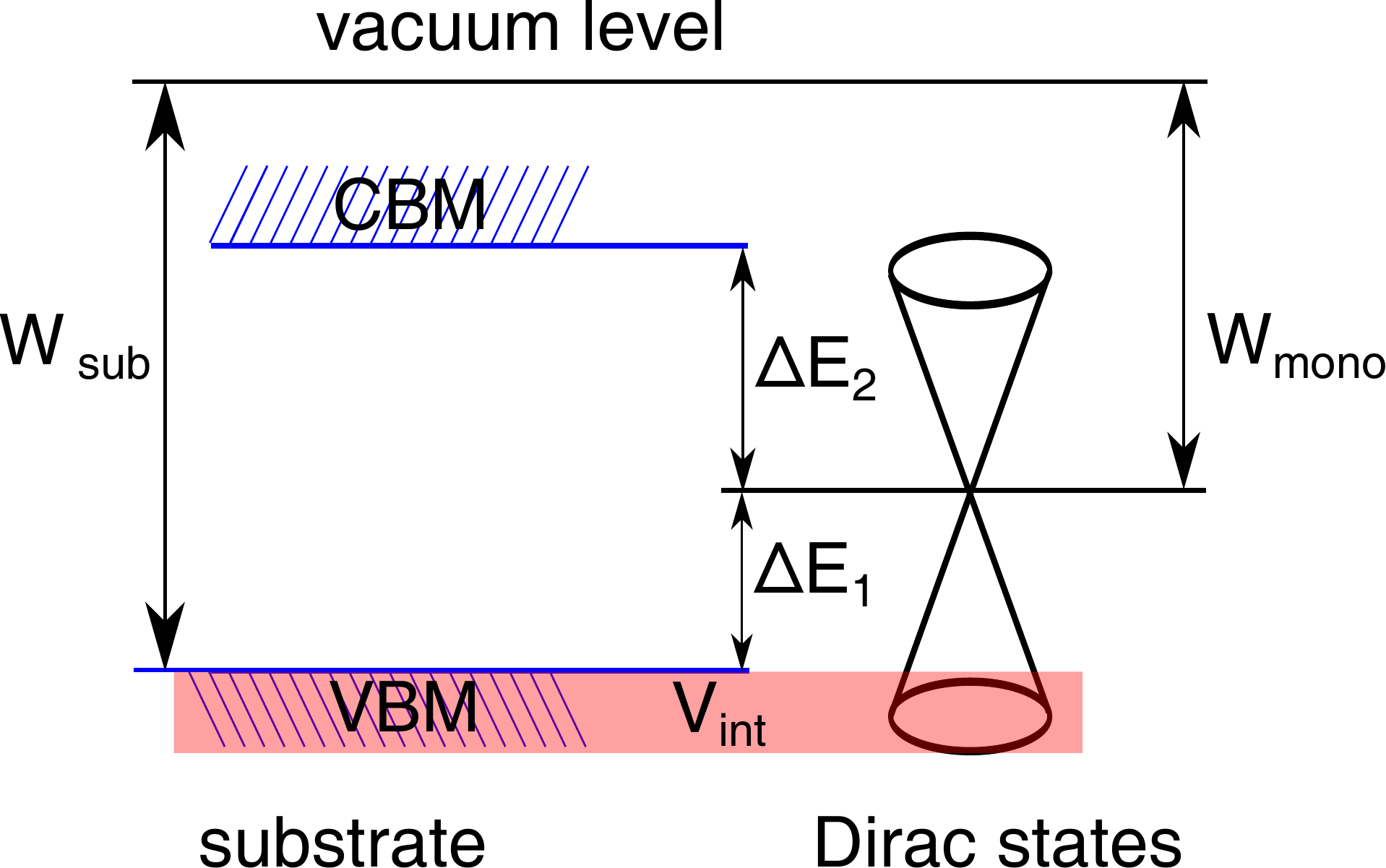}
  \caption{Schematic illustration of the interaction between a semiconducting substrate and a monolayer with Dirac states. 
  W$_{sub}$ (W$_{mono}$) represents the workfunction of the substrate (the monolayer).
  $\Delta$E$_i$ denote the energy differences between the substrate states and the Dirac point.
  V$_{int}$ stands for the interaction between the substrate and the monolayer.
  VBM and CBM represent the valence band maximum and conduction band minimum, respectively.
  }
 \label{fig1}
\end{figure}

 Our concept is illustrated in Fig.~\ref{fig1}. 
When a semiconductor substrate and a monolayer with Dirac states are
brought together, their bands are aligned to a common vacuum level.
The interaction of between them is denoted by V$_{int}$.
Similar to the heteronuclear diatomic systems 
the resulting perturbations to the electronic bands of the isolated systems are determined by both the interaction 
and the difference ($\Delta$E) in the energy levels between the two constituents.
Strong perturbations to the Dirac states can be expected when V$_{int}$ is large and $\Delta$E is small. 
In particular, the silicene/germanene bands are strongly destroyed 
when it is placed onto metal substrates 
because of the strong layer interaction and that $\Delta$E is zero.
Whereas, the Dirac states can be preserved well in vdW heterostructures since V$_{int}$ is small.
However, to stabilize the monolayer honeycomb structure
a strong binding between silicene (germanene) and the substrate is required, 
that is, V$_{int}$ is large.
Therefore, to avoid a strong disturbance to the Dirac states $\Delta$E should be large, 
that is, the Dirac states should be far from the substrate states.
For such a purpose, large-gap semiconductors with a proper workfunction are desirable 
such that the Dirac states lie in the middle of the gap when their bands align.
Of course, the interaction with the substrate may affect the structure 
and correspondingly modifies the electronic bands of the overlayer.

As a large-gap semiconductor, Al$_2$O$_3$ has been widely used as a substrate in silicon-based semiconducting devices.
A simple density-functional theory (DFT) calculation indicates that Al-terminated Al$_2$O$_3$(0001) 
is a semiconductor with a gap about 4.84 eV and a work function of about 6.60 eV.
While the ideal low-buckled silicene has a workfunction of about 4.76 eV.
By the band alignment, one can find that 
the Dirac cone of the silicene lies in the gap of Al-terminated Al$_2$O$_3$(0001), 
and is 1.94 ev and 2.90 eV from the VBM and CBM, respectively.
Such energy separations imply that the Dirac states in silicene may experience a relatively weak perturbation 
(compared to silicene/Ag(111)) even though the monolayer has a strong binding with the substrate as well.

\begin{figure}
  \includegraphics[width=0.48\textwidth]{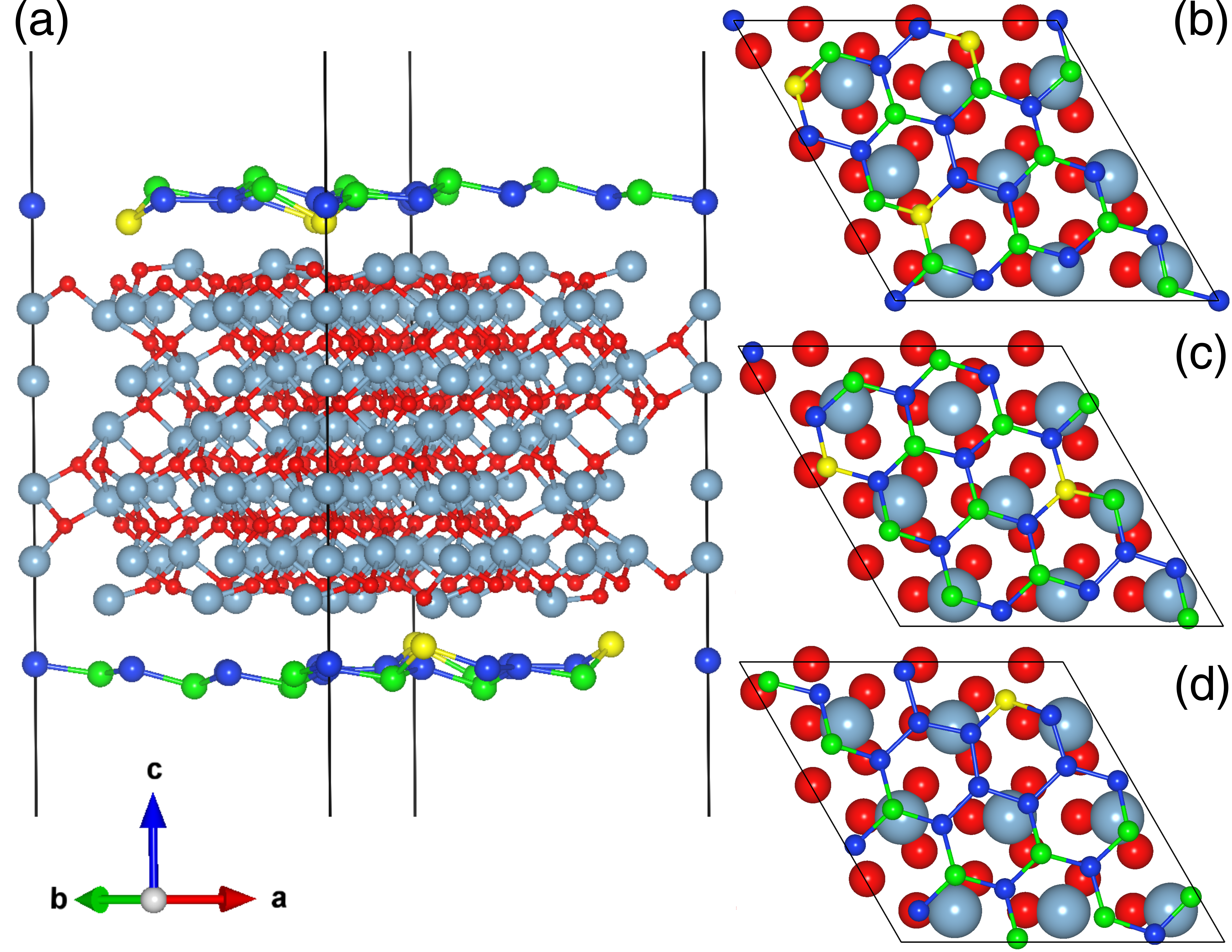}
  \caption{Structure of ($\sqrt{13}\times\sqrt{13}$) silicene on (3$\times$3) Al-terminated Al$_2$O$_3$(0001).
(a) Perspective view and (b) top view of the relaxed S1 structure.
Top view of the relaxed (c) S2 and (d) S3 structures.
Si atoms are represented by blue, green (buckling away from the substrate), and yellow (buckling towards) balls.
  }
 \label{fig2}
\end{figure}

\begin{figure*}
  \includegraphics[width=0.98\textwidth]{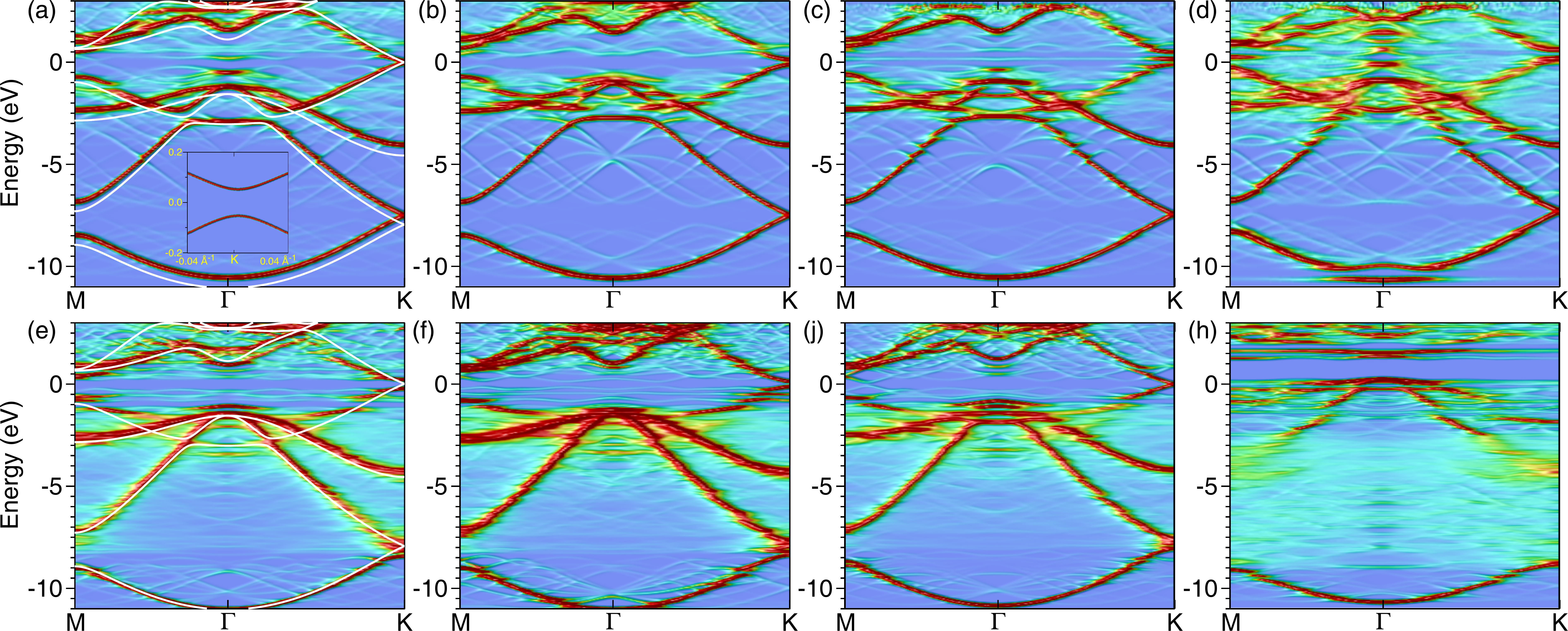}
  \caption{Electronic bands of silicene.
  (a)-(c) $k$-projected bands in the (1$\times$1) BZ for isolated silicene in the S1, S2, and S3 structures, respectively;
  (e)-(j) the corresponding bands for supported silicene on Al-terminated Al$_2$O$_3$(0001). 
  (d) and (h) $k$-projected bands for isolated and supported silicene on O-terminated Al$_2$O$_3$(0001), respectively.
  Bands of ideal(1$\times$1) silicene are shown as solid white lines overlaid for comparison.
  The inset in (a) shows bands about K in the range $\pm$0.04 \AA$^{-1}$.
  The Fermi level is set to be zero.
  }
 \label{fig3}
\end{figure*}

To validate our ideal, we performed DFT calculations for silicene on Al$_2$O$_3$(0001)
using the Vienna Ab initio Simulation Package
\cite{kresse_efficiency_1996,kresse_efficient_1996}.
The pseudopotentials were constructed by the projector augmented wave method\cite{bloechl_projector_1994,kresse_ultrasoft_1999}.
Van der Waals dispersion forces between the adsorbate and the substrate 
were accounted for through the optPBE-vdW functional by using the vdW-DF method 
developed by Klime\v{s} and Michaelides\cite{PhysRevB.83.195131,klimes_chemical_2010}.
A 4$\times$4 Monkhorst-Pack $k$-mesh was used to sample the surface BZ and 
a plane-wave energy cut off of 400 eV was used for structural relaxation and electronic structure calculations.

In our calculations the substrate Al$_2$O$_3$(0001) was modeled by a slab consisting of 6 Al$_2$O$_3$ layers (12 Al and six oxygen layers),
with the two surfaces terminated by Al-I 
since this termination was found to be more stable than the Al-II and the O-terminated surfaces\cite{Kurita_2010}. 
Because the spacing between the top Al layer and the next O layer 
is significantly reduced upon structural relaxation\cite{Kurita_2010},
the slab was firstly fully relaxed.
Then a ($\sqrt{13}\times\sqrt{13}$) silicene layer is placed on each side of a (3$\times$3) supercell of Al$_2$O$_3$(0001),
which gives rise to a small lattice mismatch ($\sim$2.8\%).
The silicene/Al$_2$O$_3$(0001) slab is separated from its periodic images by $\sim$20 \AA vacuum regions.
For the structural relaxation of silicene/Al$_2$O$_3$(0001) the Si atoms and the top Al and O atoms were allowed to relax, 
with a threshold of 0.001 eV/\AA{} for the residual force on each atom, while the positions of the other atoms were frozen.

Three types of configurations are considered for silicene/Al$_2$O$_3$(0001) as shown in Fig.~\ref{fig2}, 
which are denoted by S1, S2, and S3, respectively.
S2 (S3) has a Si atom at a top site over Al (O), 
whereas in S1 none of Si atoms is exactly sitting on a top site.
S1 is energetically more favorable than S2 and S3 by 0.023 eV/Si and 0.024 eV/Si, respectively.
Our calculations indicate that silicene - substrate distance is increased by about 0.28 \AA{} 
when vdW correction is taken into account.  
The relaxed structures of silicene are buckled similarly to ideal low-buckled silicene, but with significant deviations from the alternating
buckling pattern of ideal silicene caused by interactions with the substrate.
For S1 the large downward bucklings are about 0.9 \AA, which reduces the layer distance down to about 1.4 \AA.
Together with the low-buckled Si atoms, this gives rise to a thickness of about 1.5 \AA for the silicene. 
Si-Si bond lengths for the relaxed structures are between 2.30 and 2.42 \AA, 
comparable to those of ideal silicene (2.30 \AA). 
The binding energy of 0.52 eV/Si is comparable to 0.698 eV/Si for silicene/Ag(111)\cite{guo_absence_2013}
but much higher than that for silicene/graphene ($\sim$ 0.1 eV/Si), favoring stabilizing the monolayer structure.

The coexistence of low- and high-buckling in the supported silicene results in a reconstructed structure
which no longer has the (1$\times$1) silicene periodicity, but still maintains the basic honeycomb
structure.  Therefore, the bands can still be unfolded into the (1$\times$1) silicene BZ by projecting
the supercell wave functions onto the corresponding $k$ of the (1$\times$1) silicene cell
 \cite{bufferlayer,chen_revealing_2014}, i.e.,
the structural distortion/reconstruction serves as a perturbation to the (1$\times$1) silicene bands. 

Figures.~\ref{fig3}(a)-(c) show the $k$-projected bands for isolated silicene 
corresponding to the three configurations shown in Figs.~\ref{fig2}(b)-(d), respectively, with the bands
of the ideal silicene shown for comparison. The isolated (reconstructed) silicene bands basically
preserve the main feature of the ideal structure, although there are distinct differences.
The most prominent one is the opening of a gap at the Dirac point at K due to
the symmetry breaking caused by the structural reconstruction.
The size of the gap varies for the different structures, and is about 0.1 eV for the lowest energy structure S1. 
Moreover, the reconstruction gives rise to minigaps in the linear dispersion around K, and
slightly renormalizes the band widths compared to those of the ideal structure, c.f., Fig.~\ref{fig3}(a). 

Although the bands for silicene on Al-terminated Al$_2$O$_3$(0001),
Figs. 3(e)-3(g), are basically preserved, there are noticeable changes
 around the Fermi energy: The gap at K for the lowest energy S1 structure is enhanced to 0.44 eV; additional
minigaps are introduced (due to the different substrate periodicity); and (comparing Figs.~\ref{fig3}(a)
and (e)) the substrate induces a shift of the silicene bands of about 0.5 eV relative to the Fermi level.
In contrast, a comparison of Figs.~\ref{fig3}(d) and (h) shows that silicene bands are significantly disturbed 
by O-terminated Al$_2$O$_3$(0001).
In particular, the bands near the Fermi level lose features of gapped Dirac cone near K 
as obtained for the isolated silicene.

The band structure of silicene on Al$_2$O$_3$(0001) shows also distinct differences from metal-supported silicene
\cite{lin_substrate-induced_2013,chen_revealing_2014,guo_absence_2013,wang_absence_2013,
cahangirov_electronic_2013,quhe_does_2014,mahatha_silicene_2014} 
and silicene-based vdW heterostructures
\cite{guo_absence_2013,cai_stability_2013,Houssa_APL_2010,Kaloni_JAP_2013,Ding_APL_2013,Scalise_2D_2014,Kokott_JPCM_2014,C4CP00089G}.
In the former case, the silicene bands are strongly hybridized with those of the substrate
so that the bands at the K point are significantly destroyed, no longer having Dirac-electron character, 
while the linear dispersions are well preserved in silicene-based vdW heterostructures.
Moreover, the binding between silicene and Al-terminated Al$_2$O$_3$(0001), together with  
the structural and electronic properties of the supported silicene, 
 support that silicene can be properly stabilized by a Al$_2$O$_3$ encapsulation layer\cite{ADFM:ADFM201300354}, 
although the encapsulation layer is likely amorphous in the experiment.

\begin{figure}
  \includegraphics[width=0.48\textwidth]{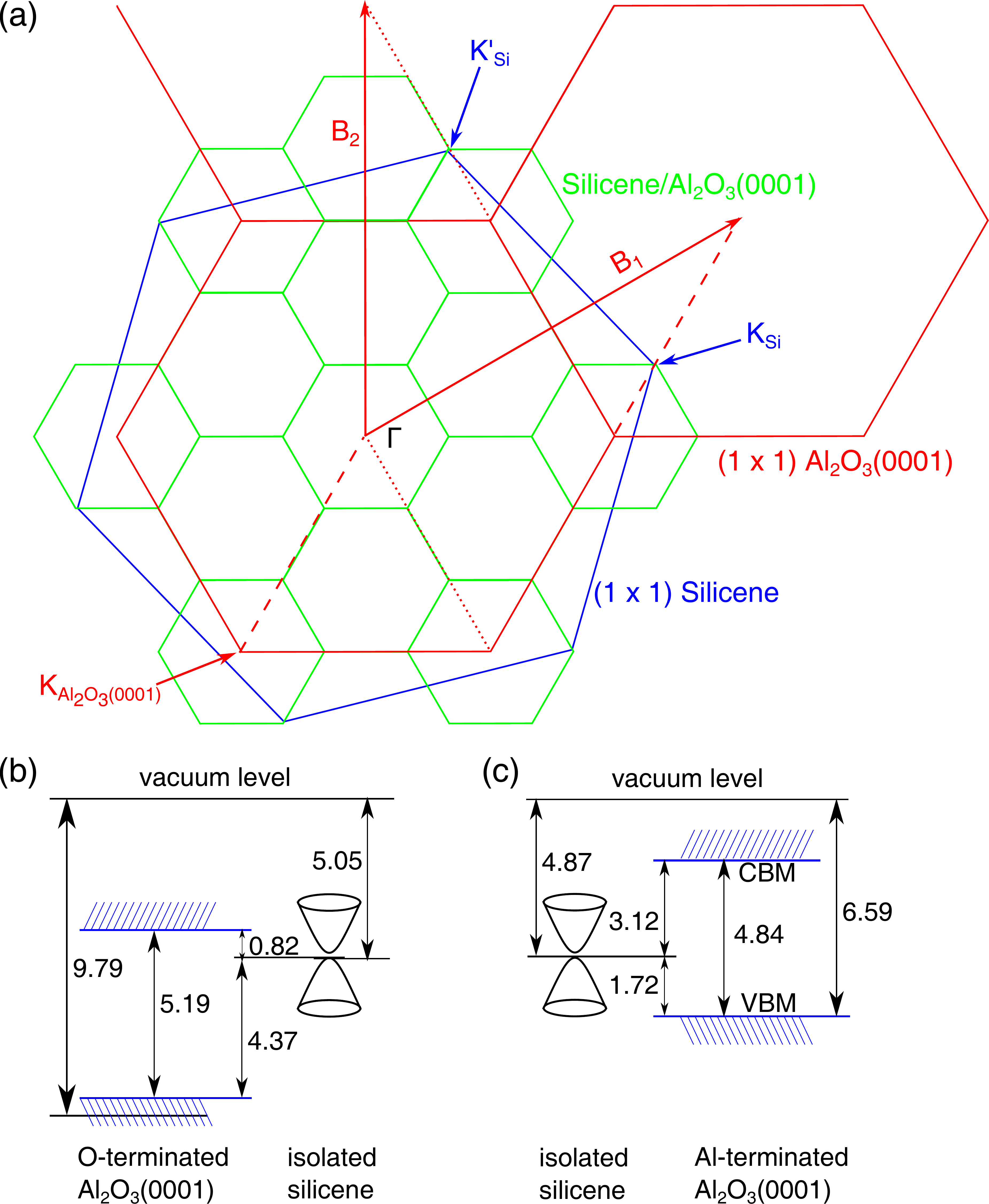}
  \caption{Interaction between silicene and Al$_2$O$_3$(0001).
(a) BZs of silicene/Al$_2$O$_3$(0001), (1$\times$1) silicene and Al$_2$O$_3$(0001) 
shown in green, blue and red, respectively.
$\mathbf{B_1}$ and $\mathbf{B_2}$ represent the reciprocal lattice vectors of (1$\times$1) Al$_2$O$_3$(0001).
(b), (c) Band alignment of the isolated silicene and  
O-terminated and Al-terminated (0001) surfaces of Al$_2$O$_3$, respectively.
In the case of Al-terminated Al$_2$O$_3$(0001), the isolated systems are obtained from configuration S1.   
The calculated work functions and the gaps of Al$_2$O$_3$(0001) are shown.
VBM and CBM represent the valence band maximum and conduction band minimum, respectively.
  }
 \label{fig4}
\end{figure}
 
To gain further insights into the interaction between the states of silicene and Al$_2$O$_3$(0001),
Fig.~\ref{fig4}(a) shows the supercell BZs of silicene/Al$_2$O$_3$(0001), and the BZs of the ideal
silicene and the substrate. Note that the 2/3K$_\mathrm{Al_2O_3}$ (K$_\mathrm{Al_2O_3}'$) coincides with
K$_\mathrm{Si}$ (K$_\mathrm{Si}'$) by a translation of a reciprocal lattice vector of (1$\times$1)
Al$_2$O$_3$(0001), $\mathbf{B_1}$ ($\mathbf{B_2}$).
When silicene is overlaid onto Al$_2$O$_3$(0001),  
the wavefunctions of silicene $\Psi_\mathrm{Si}$ ($K_\mathrm{Si}$) interact with those of Al$_2$O$_3$(0001) 
$\Psi_\mathrm{Al_2O_3}$ ($\frac{2}{3}$ $K_\mathrm{Al_2O_3}$+$\mathbf{B_1}$).
Figs.~\ref{fig4}(b) and (c) shows the band alignment of the two constituents before interacting.
In the case of silicene/O-terminated-Al$_2$O$_3$(0001)
the gapped Dirac states of silicene are close to the conduction band of the substrate.
This along with the strong interaction between them (E$_b$ is larger than 1.0 eV) 
lead to strong modifications over the silicene bands.
Whereas, for silicene/Al-terminated-Al$_2$O$_3$(0001) (configuration S1)
the Dirac states of silicene are in the gap of the substrate,
far from the substrate states.
Therefore despite the strong binding between the two constituents, 
there is little direct bonding of between the silicene Dirac states and the Al$_2$O$_3$ states.
By projecting the wavefunctions over atomic orbitals, 
we find that Si-pz orbitals contribute over 75\% to the VBM and CBM. 
Whereas, the substrate just has little contributions.

\begin{figure}
  \includegraphics[width=0.48\textwidth]{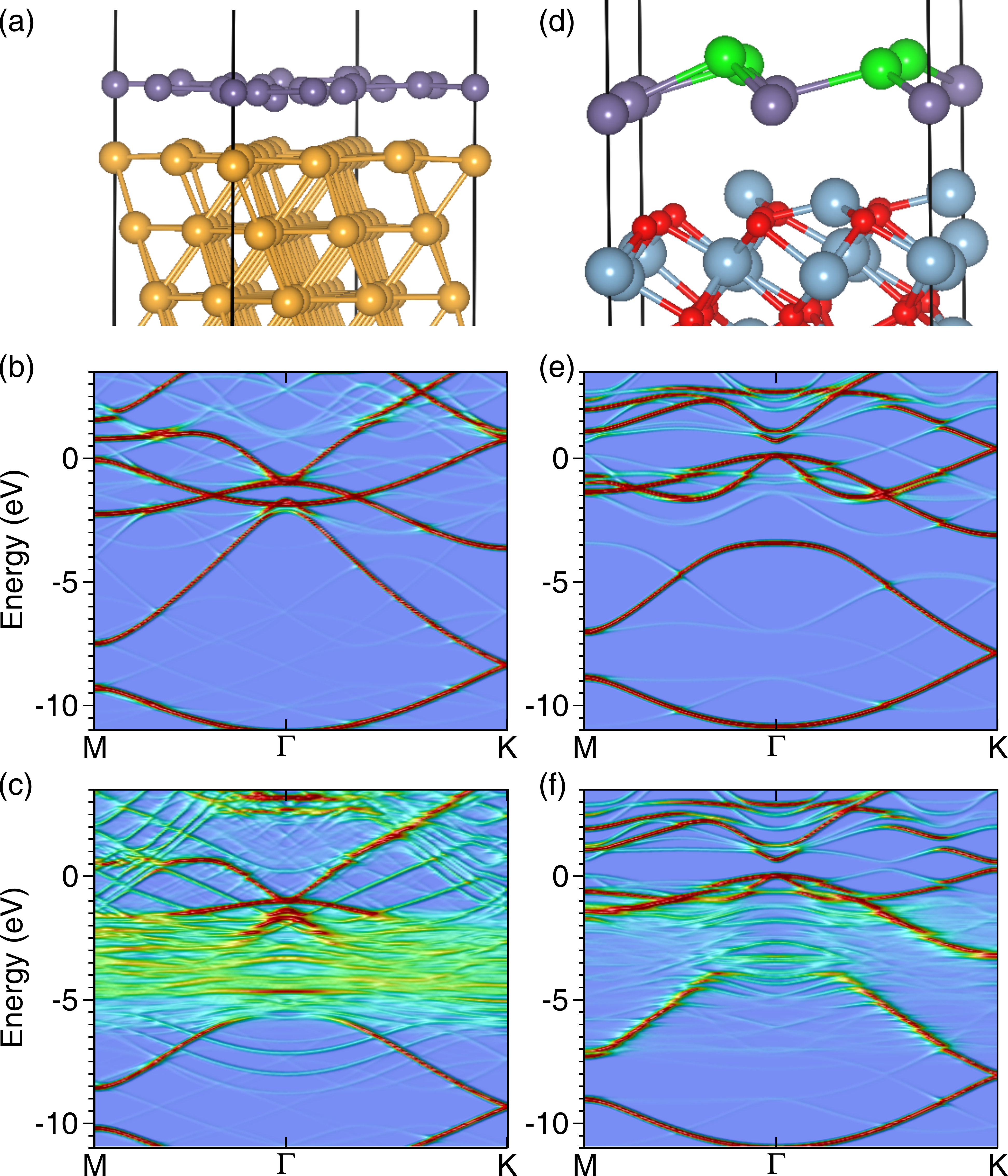}
  \caption{DFT calculations of germanene on Au(111) and Al$_2$O$_3$(0001). 
  (a) Relaxed structure of (3$\times$3) germanene on ($\sqrt{19}\times\sqrt{19}$) Au(111).
  (b) and (c) $k$-projected bands for isolated and supported germanene, respectively.
  (d) - (f) Corresponding results for (2$\times$2) germanene on ($\sqrt{3}\times\sqrt{3}$) Al$_2$O$_3$(0001).
  Buckled Ge atoms are shown in green balls.
  }
 \label{fig5}
\end{figure}
Likewise, a DFT calculation shows that the work function of an ideal low-buckled germanene is about 4.66 eV,
which will place the Dirac point in the gap of Al-terminated Al$_2$O$_3$(0001) according to the band alignment and
will gives rise to an energy separation of about 1.93 (2.91) eV between the Dirac point and the VBM (CBM) of the substrate, respectively.
Therefore, one may expect that the Dirac states are likely preserved when the germanene monolayer is overlaid onto 
Al-terminated Al$_2$O$_3$(0001).
DFT calculations were then performed for germanene on Al-terminated Al$_2$O$_3$(0001), 
of which the $k$-projected bands are shown in Fig.~\ref{fig5}.
The band structures for germanene on Au(111) are also shown for comparison.
The relaxed structure of germanene on Au(111) is pretty much similar to the planar hexagonal structure.
Therefore, the $k$-projected bands for the isolated germanene are in great similarity to those for the planar germanene
\cite{cahangirov_two-_2009}.   
However, its electronic bands experience dramatic changes when germanene is placed on Au(111).
From Fig.~\ref{fig5}(c) one can see that germanene-derived bands in the energy window of -6.0 eV to 3.0 eV are strongly destroyed.
In particular, near the Fermi level the linear dispersions at the K point become indistinguishable.
This is due to the strong hybridization between the substrate bands and the Dirac states. 
However, germanene on Al$_2$O$_3$ exhibits distinct differences. 
Fig.~\ref{fig5}(d) shows that the relaxed structure of supported germanene 
basically remains the global profile of the ideal low-buckled structure.
Therefore it is not surprising that the $k$-projected band structure for the isolated germanene 
bears a great similarity to that for the ideal germanene, except for minigaps near $\pm$ 1.0 eV (Fig.~\ref{fig5}(e)).
Upon the presence of Al-terminated Al$_2$O$_3$(0001), 
the linear dispersions near K are preserved well, except for the minigap near 1.0 eV and the gap at K are slightly enhanced.

It is demonstrated that external electric field can be used to tune the energy gap of silicene
\cite{drummond_electrically_2012,Ezawa_NJP_2012}.
Therefore, in addition to band hybridization, 
electric field induced by the substrate may be responsible for the changes 
in the band gaps of silicene and germanene upon supported.
It should be mentioned that the energy gaps of our systems obtained using the standard DFT may be underestimated.
More accurate descriptions can be achieved by using hybrid functionals and GW methods.
However, such calculations are not feasible for our systems with a larger number of atoms in the unit cell.
Nevertheless, our results indicate that the Al-terminated Al$_2$O$_3$(0001) 
not only stabilizes the monolayer structure of silicene and germanene,
but also tunes their energy gap for potential applications in electronic devices.

\begin{figure}
  \includegraphics[width=0.40\textwidth]{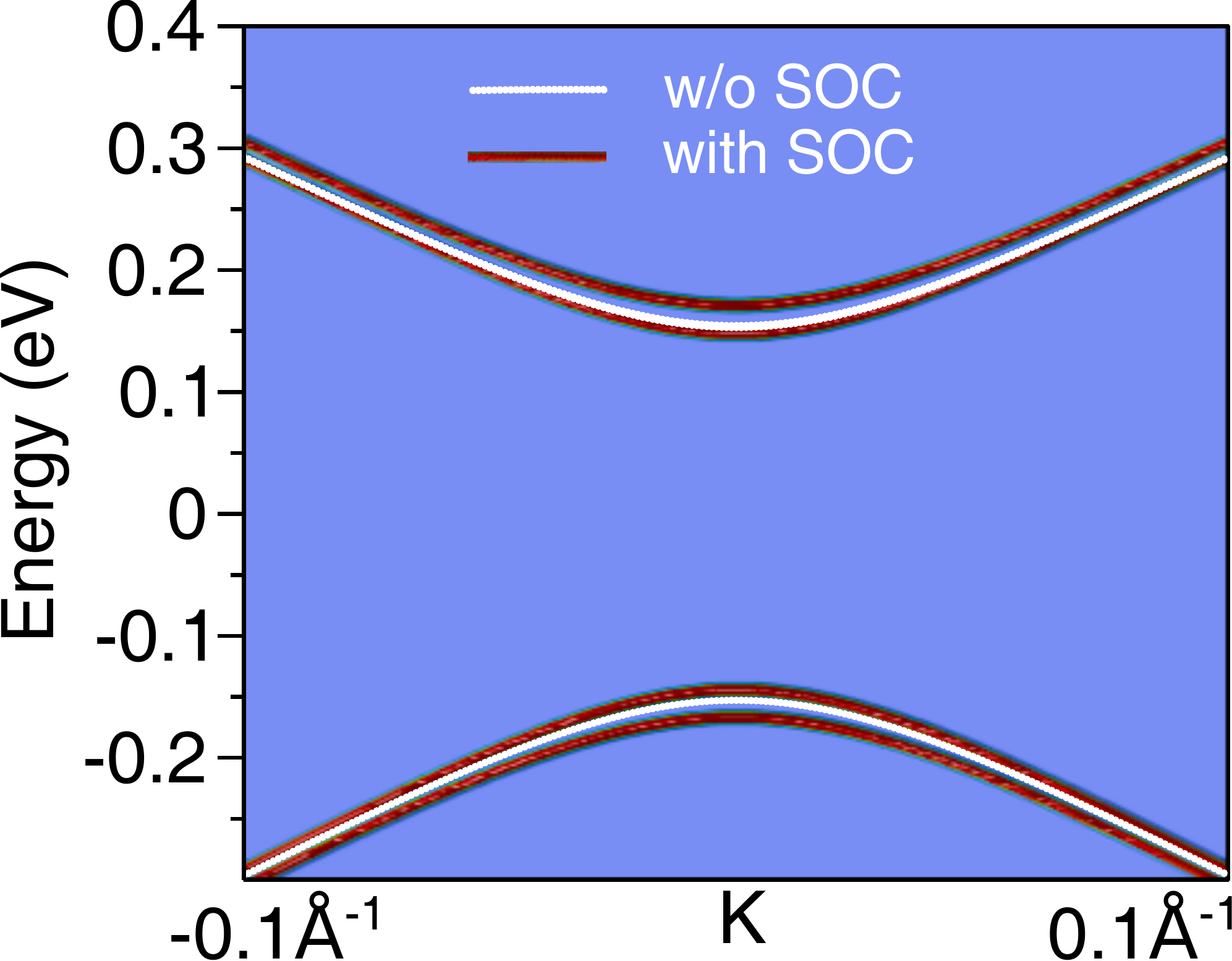}
  \caption{Band structure with SOC about the K point for germanene/Al$_2$O$_3$(0001). 
  The bands without SOC (in white) are overlaid for comparison.
  }
 \label{fig6}
\end{figure}
Spin-orbit coupling (SOC) is important for the predicted quantum spin Hall effect in the ideal low-buckled silicene and germanene,
which induces a gap opening at the K point.
The SOC gaps are about 1.55 meV and 23.9 meV for silicene and germanene, respectively\cite{liu_quantum_2011}. 
Our calculations without including SOC show that gap opening occurs in the presence of the substrate 
(see Fig.~\ref{fig3} and Fig.~\ref{fig5}), which is due to the symmetry breaking in the sublattice.
SOC calculations were further performed to investigate the effect of SOC on the band structure of germanene/Al$_2$O$_3$(0001).
Fig.~\ref{fig6} shows that SOC induces splittings in both the valence and the conduction bands,
which are about 25 meV at the K point.
Likewise, our calculations (plot not shown) for the case of silicene find that SOC splittings at K are about 1.5 meV.

\begin{figure}
  \includegraphics[width=0.48\textwidth]{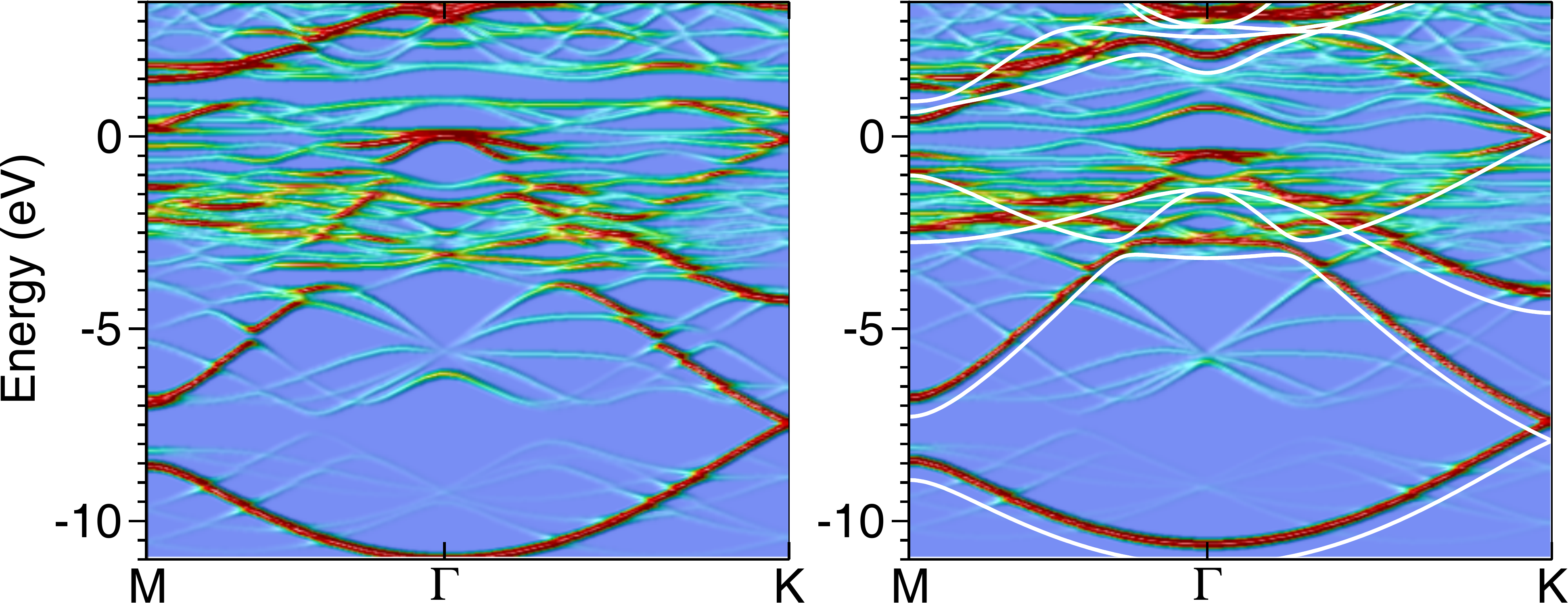}
  \caption{$k$-projected bands for isolated silicene in  the structure calculated for  
  (a) ($\sqrt{12} \times \sqrt{12}$) and (b) ($\sqrt{13} \times \sqrt{13}$)  Ag(111).
  Bands for ideal silicene are shown as white lines for comparison. 
  }
 \label{fig7}
\end{figure}
Our results can shed light on the degradation of carrier mobility observed for silicene-field effect transistors, 
where silicenes directly contact with Al$_2$O$_3$ (although the substrate may be amorphous), that is, 
the measured electron mobility was an order of magnitude lower than predicted for the ideal silicene
\cite{tao_silicene_2015}.
An important issue is that the silicenes investigated in the present study differ from the experimental ones.
In Tao's experiments, silicene is either in ($\sqrt{12}\times\sqrt{12}$) or ($\sqrt{13}\times\sqrt{13}$)
with respect to the ($1\times1$) Ag(111) (which is ($\sqrt{7}\times\sqrt{7}$) with respect to the
(1$\times$1) silicene).
Fig.~\ref{fig7} shows $k$-projected bands for isolated silicene in the structures corresponding to 
 ($\sqrt{12}\times\sqrt{12}$) and ($\sqrt{13}\times\sqrt{13}$) Ag(111).
The folded bands for these structures agree well those reported by previous studies\cite{guo_absence_2013,scalise_vibrational_2014}.
Although Dirac-like bands at K are observed in the folded band structure for the isolated silicene on
($\sqrt{13}\times\sqrt{13}$) Ag(111),
$k$-projected bands indicate that silicene bands undergo considerable changes upon the structural distortion.
The effects of the structural reconstruction are much more pronounced for the supported silicene 
on ($\sqrt{12} \times \sqrt{12}$)  Ag(111).
For ($\sqrt{7}\times\sqrt{7}$) silicene  placed onto Al$_2$O$_3$(0001), a supercell of
($\sqrt{21}\times\sqrt{21}$) is required to obtain a small lattice mismatch of $\sim$2\%, and a
corresponding substrate ($\sqrt{13}\times\sqrt{13}$) supercell. While such calculations might be
possible, we have not attempted them; nevertheless, based on the above results one may expect similar
effects of the substrate on the ($\sqrt{7}\times\sqrt{7}$) silicene as seen here.  Moreover, our
calculations are consistent with experiments that a gap is formed in the band structure of silicene on
Al$_2$O$_3$ (Fig.~\ref{fig3}).
  
The electron mobility is determined by the Fermi velocity and 
relaxation times of electrons due to various scattering processes, i.e., $\mu = v^2\tau$.
Fig.~\ref{fig3} shows that the structural reconstruction renormalizes the electronic bands 
and therefore renormalizes the Fermi velocity as well.
By a comparison of bands for the ideal silicene and the isolated one (Fig.~\ref{fig3}(a)),
a decrease in the Fermi velocity for the supported silicene is found.
Moreover, the substrate interactions and the silicene reconstruction induce minigaps in the band structure of silicene, 
which also significantly reduce the Fermi velocity near the minigaps.
The structural reconstructions also modify the phonon spectrum of silicene and thus the intrinsic electron-phonon
scattering relaxation time.
In addition, surface polar phonon of the substrate may play a role 
in scatterings electrons in the overlayer
\cite{substrate-limited_2008,chen_intrinsic_2008}, further reducing the electron mobility.

In summary, we propose a guideline for exploring suitable substrates for silicene and germanene.
Our DFT calculations find that Al-terminated Al$_2$O$_3$(0001) may be a good candidate:
although silicene (germanene) experiences structural distortions due to the interaction with the substrate, 
the main profile of the ideal low-buckled structure is maintained.
Unlike the metal-supported monolayers, where Dirac electrons are absent,
a gapped Dirac cone is formed at the K point in silicene as well as germanene on Al-terminated Al$_2$O$_3$(0001).
However, Al$_2$O$_3$ as a substrate has substantial effects on the electronic bands of silicene:
not only inducing a band gap at the K point, but also giving rise to minigaps in the band structure, 
which will lead to Fermi velocity renormalization and thus affect the electronic transport properties.
Our results not only can shed light on recent transport experiments on silicene-based transistors,
but also may aid in the design of silicene/germanene-based electronic devices.

Note added in proof: Since submission of this paper,
germanene was grown on defected MoS2,
where germanene islands preferentially nucleate at the defects.\cite{zhang_structural_2016}

\begin{acknowledgments}   
This material is based upon work supported by the: U.S. Department
    of Energy, Office of Science, Office of Basic Energy Sciences, under
    Award No. DE-FG02-05ER46228.
\end {acknowledgments}

\bibliography{references}
\bibliographystyle{apsrev4-1}
\end{document}